\documentclass[fleqn,usenatbib]{mnras}
\usepackage{newtxtext,newtxmath}
\usepackage{multirow}
\usepackage{epsfig}
\usepackage{amsmath,color,bm}
\usepackage{booktabs}
\usepackage{subfig}

\makeatletter
\renewcommand{\p@subfigure}{}
\makeatother
\usepackage{nameref}
\usepackage{graphicx}
\usepackage{graphics}
\usepackage[space]{grffile}
\usepackage{dblfloatfix}
\usepackage{url}
\usepackage[utf8]{inputenc}
\usepackage{fancyref}
\usepackage{xcolor}
\usepackage[normalem]{ulem}
\usepackage{eqnarray,amsmath}
\usepackage{multicol}

\definecolor{magenta}{rgb}{139, 0, 139}

\title[Improved Early-Warning with Higher Modes]{Improved early-warning estimates of luminosity distance and orbital inclination of compact binary mergers using higher modes of gravitational radiation}

\author[M. K. Singh et al.]{Mukesh Kumar Singh,$^1$
  Divyajyoti,$^{2, 3}$
  Shasvath J. Kapadia,$^1$
  Md Arif Shaikh,$^1$ \newauthor
  and Parameswaran Ajith$^{1,4}$\\
  $^1$International Centre for Theoretical Sciences, Tata Institute of Fundamental Research, Bangalore 560089, India\\
  $^2$Indian Institute of Technology Madras, Chennai 600036, India\\
  $^3$Centre for Strings, Gravitation and Cosmology, Department of Physics, Indian Institute of Technology Madras, Chennai 600036, India \\
  $^4$Canadian Institute for Advanced Research, CIFAR Azrieli Global Scholar, MaRS Centre, West Tower, 661 University Ave, Toronto, ON M5G 1M1, Canada}

\begin{document}
\label{firstpage}
\pagerange{\pageref{firstpage}--\pageref{lastpage}}
\maketitle

\begin{abstract}
The pre-merger (early-warning) gravitational-wave (GW) detection and localization of a compact binary merger would enable astronomers to capture potential electromagnetic (EM) emissions around the time of the merger, thus shedding light on the complex physics of the merger. While early detection and sky localization are of primary importance to the multimessenger follow-up of the event, improved estimates of luminosity distance and orbital inclination could also provide insights on the observability of the EM emission. In this work, we demonstrate that the inclusion of higher modes of gravitational radiation, which vibrate at higher multiples of the orbital frequency than the dominant mode, would significantly improve the early-warning estimates of the luminosity distance and orbital inclination of the binary. This will help astronomers to better determine their follow-up strategy. Focusing on future observing runs of the ground-based GW detector network [O5 run of LIGO-Virgo-KAGRA, Voyager, and third-generation (3G) detectors], we show that for a range of masses spanning the neutron-star black-hole binaries that are potentially EM-bright, the inclusion of higher modes improve the luminosity distance estimates by a factor of $\sim 1-1.5 \ (1.1-2) \ [1.1-5]$ for the O5 (Voyager) [3G] observing scenario, 45 (45) [300] seconds before the merger for the sources located at 100 Mpc. There are significant improvements in orbital inclination estimates as well. We also investigate these improvements with varying sky-location and polarization angle. Combining the luminosity distance uncertainties with localization skyarea estimates, we find that the number of galaxies within localization volume is reduced by a factor of $\sim 1-2.5 \ (1.2-4) \ [1.2-10]$ with the inclusion of higher modes at early-warning time of 45  (45) [300] seconds in O5 (Voyager) [3G].   
\end{abstract}

\section{Introduction}\label{sec:introduction}
	The LIGO-Virgo \citep{advligo, advvirgo} network of ground-based interferometric detectors has completed three observing runs so far. These have provided over $\sim 90$ gravitational-wave (GW) detections \citep{gwtc-1, gwtc-2, gwtc-3, ias-4, Nitz_OGC4}. Most of the GWs are found to be consistent with binary black hole (BBH) coalescences, although neutron-star black-hole (NSBH) and binary neutron star (BNS) mergers have also been observed \citep{gwtc-1, gwtc-2, gwtc-3, ias-1, ias-2, ias-3, ias-4, Nitz_OGC1, Nitz_OGC2, Nitz_OGC3, Nitz_OGC4}.

The very first BNS merger detected by LIGO-Virgo --- GW170817 \citep{GW170817-DETECTION} --- was accompanied by an electromagnetic (EM) counterpart that was followed up by numerous telescopes worldwide \citep{GW170817-MMA}. The resulting scientific gain was spectacular: it enabled an unprecedented probe of the neutron star equation of state (EOS) \citep{GW170817-EOS},  a stringent constraint on the deviation of the speed of GWs with respect to the speed of light \citep{Liu2020} and alternative theories of gravity \citep{GW170817-TGR}, a distance ladder independent measurement of the Hubble constant \citep{GW170817-HUBBLE}, and ascertained BNS mergers to be sites where heavy elements of the periodic table get synthesized \citep{GW170817HeavyElements}.

Future observing runs could potentially detect compact binary coalescences (CBCs) with EM-counterparts, although the relative rarity of such events motivates the need to maximize the science gains afforded by them. In particular, a pre-merger/early-warning (EW) detection and localization of these events could allow telescopes to capture potential precursors \citep[e.g.][]{InspiralCounterpart} and signatures of intermediate products such as hypermassive or supramassive neutron stars \citep[e.g.][]{HotokezakaHNS}. 

Current EW efforts are targeted towards BNSs \citep{SachdevEW, MageeEW}, exploiting the relatively longer duration of their GWs in the band of ground-based detectors as compared to BBH mergers.  This allows templated matched-filter searches to accummulate sufficient signal-to-noise ratio (SNR) several seconds to a minute before merger \citep{CannonEW}, thus enabling a pre-merger detection and localization.

Certain NSBH mergers are also expected to produce EM-counterparts \footnote{We refer to mergers with EM counterparts as ``EM-bright''}, depending on the mass ratio of the binary, the spin of the BH, and the EOS of the NS \citep{FoucartEMB1, FoucartEMB2}. However, their heavier total mass compared to BNSs reduce their in-band duration \citep{Sathyaprakash1994}, thus making an EW detection and localization more challenging. 

Current real-time (low-latency) CBC searches \citep{gstlal, mbta, pycbc,spiir} use GW templates that consist of only the dominant harmonic. On the other-hand, for asymmetric mass systems such as NSBHs with moderate to high orbital inclination, the subdominant harmonics can contribute a significant fraction of the SNR \citep[e.g.,][]{Varma2014}. Recent work showed that the inclusion of subdominant modes in low-latency searches would improve EW detection and sky-localisation, by virtue of the fact that these modes enter the frequency band of the detectors well before the dominant mode\footnote{In \cite{nsbh_EW2, nsbh_EW1}, the effect of precession has also been explored as a potential way to improve the early warning efforts of NSBH binary mergers.} \citep{Kapadia2020, Singh_EW2021}.

Not only the sky-localization but also the measurement of luminosity distance and orbital inclination, in the early-warning time, can be improved significantly with the inclusion of higher modes in online GW searches. The increased information content in the higher modes can also help in breaking degeneracies between the luminosity distance and orbital inclination. The improved measurements of luminosity distance will help in better understanding if an EM emission from the source can be observed by the existing ground/space based telescopes. Similarly, the better estimates of inclination angle will help in informing about the observability of any beamed EM emission, for example, short off-axis gamma ray bursts (GRBs) from the source \citep{arun_sGRB}. Thus, the improved measurements of both of these quantities will help astronomers to decide their follow-up strategies accordingly.

In this work, we demonstrate the benefits of including subdmoniant modes in real-time searches, in the context of EW. Specifically, using a Fisher Matrix based analysis, we show that estimates of luminosity distance and orbital inclination, improve considerably in EW time, with the inclusion of higher modes. We consider three observing scenarios: O5, Voyager, and 3G. ``O5'' consists of the LIGO-Virgo-Kagra network, including LIGO-India \citep{LIGO-INDIA, SaleemLIGOIndia}, operating at their A+ \citep{aplus_sensitivity} sensitivities.  In the Voyager \citep{Voyager_PSD, Adhikari_voy} scenario, the three LIGO detectors have their sensitivities upgraded. The 3G network has two Cosmic Explorers, \citep{CE} and one Einstein Telescope \citep{ET}. We find that the inclusion of higher modes reduces the error uncertainties on the luminosity distance estimates of potentially EM-Bright \citep{ChatterjeeEMB} NSBH binaries located at $100$ Mpc by a factor of $\sim 1-1.5 (1.1-2)[1.1-5]$ in the O5 (Voyager) [3G] scenario, $45 (45) [300]$ seconds before merger. Combining these uncertainties with sky area estimates, and assuming a galaxy number density of $0.01 \mathrm{Mpc}^{-3}$ \citep{galaxy_density}, we find that the number of galaxies within the uncertainty volume is reduced by a factor of $\sim 1-2.5(1.2-4)[1.2-10]$~\footnote{The factors of improvements in the measurement of various parameters with the inclusion of higher modes quoted in this work are more realistic in comparison to the factors of improvements in skyarea-localization in \cite{Kapadia2020} and \cite{Singh_EW2021}. These works overestimate the improvement-factors since they do not account for the effect of priors.}.

The rest of the paper is organized as follows. Section~\ref{sec:method} explains how the inclusion of higher modes improves EW estimates of extrinsic parameters.  It further summarizes the Fisher Matrix/quadratic approximation to the GW parameter estimation likelihood. It also describes the key error-uncertainty formulae as well as the multipole expansion of the GW strain.  Section ~\ref{sec:results} outlines the results which demonstrate the benefits of the inclusion of higher modes in early-warning efforts targeted at NSBH systems. The paper concludes with Section ~\ref{sec:conclusion} where the EW gains afforded by higher modes are discussed in the context of EM follow-up.

\section{Motivation and Method}\label{sec:method}
\subsection{Higher order modes and early warning}
The GW strain can be expressed as a complex combination of its two polarizations, `$+$" and ``$\times$", as $h(t):= h_+(t) - i h_{\times}(t)$. This combination can further be expanded in basis of spin$-2$ weighted spherical harmonics \citep{NewmanPenrose}:
\begin{equation}
h(t;\theta) = \frac{1}{d_L}\sum_{\ell = 2}^{\infty} \sum_{m = -\ell}^{\ell} h_{\ell m}(t;\vec{\lambda}) Y_{-2}^{\ell m}(\iota, \varphi_0)
\end{equation} 
where $h_{\ell m}(t;\vec{\lambda})$ are called the multipoles of the radiation which depend on time $t$ as well as the intrinsic parameters of the source (such as component masses and spins in the case of a binary system). The spin weighted spherical harmonics $Y_{-2}^{\ell m}(\iota, \varphi_0)$ capture the angular dependence of the source's orientation ($\iota$) in the sky with respect to the line of sight of the observer and reference phase ($\varphi_0$) defined in terms of the rotation of the source frame with respect to the detector frame. Here, $d_L$ is the luminosity distance of the source.

Since the monopole ($\ell=0$) and the dipole ($\ell=1$) terms of the above multipolar expansion vanish due to the conservation of mass (or energy) and linear momentum of the source respectively, the dominant contribution to the GW signal comes from the quadrupole mode ($\ell=2, m=\pm 2$).  Several studies \citep[see, e.g.][]{varma_ajith} have shown that CBC templates modelled with only the dominant mode are sufficient to analyse the data containing the signal produced by compact binaries with near-symmetric masses. But the contribution of ``higher order'' (or ``subdominant'' or ``non-quadrupole'') modes becomes important for GW signals emitted by highly asymmetric sources. Also, the relative contribution of higher order modes increases with increasing asymmetry in the system, contributed by high mass ratio, spin precession, etc. The relative contribution of higher modes in the observed signal will be significant for binaries with high inclination angles. Neglecting higher modes for such systems can bias the inference of the astrophysical properties of the sources \citep{Varma2014, varma_ajith}. Higher modes will also improve the detectability of binaries as well as the precision with which the source parameters can be estimated \citep{Chris_Anand, Divya_HMs, Harry_HM_search, Capano_HM_search, imbh_hm, Arun_HMs}.

Higher order modes are not only necessary for unbiased and precise astrophysical inference of the source properties but can also improve the early-warning capabilities (localizing a source in the sky prior to the merger) of the compact binary mergers as shown in recent works \citep{Kapadia2020, Singh_EW2021}. 
Recent studies have shown that including higher order modes in the online searches can improve the search sensitivity of the detectors for asymmetric compact binary mergers \citep{Capano_HM_search, Harry_HM_search}. 

Assuming a non-precessing binary, the instantaneous frequency $f_{\ell m}(t)$ at which the higher modes vibrate is an integer multiple of the orbital frequency $f_{\mathrm{orb}}(t)$ of the compact binary:
\begin{equation}
f_{\ell m}(t) \approx m f_{\mathrm{orb}}(t)
\label{eq:multipole_f_relation}
\end{equation}
As a result, higher modes ($m>2$) enter the sensitivity band of the detector well before the dominant mode ($\ell=2,m=\pm 2$).  Thus, the in-band duration of a GW signal is effectively increased with the inclusion of subdominant modes \citep{Sathyaprakash1994}: 
\begin{equation}
\tau_c \approx \frac{5}{256} \mathcal{M}^{-5/3} (2 \pi f_{\mathrm{orb}})^{-8/3} \propto (f_{\ell m}/m)^{-8/3}
\end{equation}
where $\mathcal{M}:= (m_1 m_2)^{3/5}/(m_1 + m_2)^{1/5}$ is the chirp mass of the binary. The in-band duration of ($\ell$, $m$) mode is larger by a factor of $(m/2)^{8/3}$ as compared to the dominant mode. For example, the $\ell =3, m=3$ and $\ell =4, m=4$ modes will spend $\sim 3$ and $\sim 6$ times as long in-band as compared to the quadrupole mode.

\subsection{Parameter estimation} 
A fully Bayesian GW parameter estimation (PE) exercise to infer the parameters $\vec{\theta}$ of the binary that produced a CBC signal in the data $s$ requires the sampling of the likelihood $p(s|\vec{\theta})$ in a large-dimensional parameter space. However, this turns out to be computationally expensive and time consuming in general.  A common workaround, is to expand the log-likelihood in source parameters and truncate at quadratic order. The covariance of the resulting multidimensional Gaussian is given by inverse of the Fisher information matrix \citep{Cutler_Flanagan}. This approximation works well for high-SNR signals and breaks down at very low SNRs \citep{Vallisneri}. 

We denote by $s(t)$ the detector strain time series, which consists of noise $n(t)$,  and a GW CBC signal $h(t)$ as well:
\begin{equation}
s(t) = n(t) + h(t).
\end{equation}
Assuming that the noise is stationary and Gaussian, the likelihood on the binary's parameters $\vec{\theta}$ is given by:
\begin{equation}
p(s|\vec{\theta}) \propto e^{-(s - h(\vec{\theta})|s - h(\vec{\theta}))/2},
\end{equation}
where $(\cdot|\cdot)$ denotes the noise-weighted inner product given by:
\begin{equation}
(a|b) = 2 \int_{0}^{\infty} \frac{\tilde{a}(f) \tilde{b}^* (f) + \tilde{a}^*(f) \tilde{b}(f)}{S(f)}df.
\end{equation}
Here $\tilde{a}(f)$ and $\tilde{b}(f)$ are the Fourier transforms of $a$ and $b$ respectively,  $*$ denotes the complex conjugate, and $S(f)$ denotes the noise power spectral density (PSD).

Expanding the log-likelihood to quadratic order about the peak of the distribution yields: 
\begin{equation}
p(s\mid\vec{\theta}) \propto e^{-\frac{1}{2}\Gamma_{ij}\Delta\theta_i \Delta\theta_j},
\label{eq:err_prob}
\end{equation}
where $\Delta \theta_i \equiv \theta_i - \bar{\theta}_i$, and $\bar{\theta}_i$ corresponds to the peak of the likelihood. The quantity $\Gamma$ is the so-called Fisher information matrix and is defined for the $k^{\mathrm{th}}$ detector as,
\begin{equation}
\Gamma_{ij}^{k} = \left(\frac{\partial h^k}{\partial \theta_i} \Big \rvert \frac{\partial h^k}{\partial \theta_j} \right),
\end{equation}
The net Fisher matrix in case of a network of detectors is
\begin{equation}
    \Gamma = \sum_{k} \Gamma^{k}. 
\end{equation}
The size of the approximate likelihood function [as well as the posterior distribution $p(\theta | s)$ assuming flat priors] is given by the covariance matrix ($\Sigma$), which is related to Fisher information matrix as follows,
\begin{equation}
\Sigma_{ij} = \langle \Delta \theta_i \Delta \theta_j \rangle = (\Gamma_{ij})^{-1}
\end{equation}
We can relate the width of the 1-sigma confidence region of the posterior of parameter $\theta_i$ (marginalized over all other parameters) to the diagonal elements of the covariance matrix as,
\begin{equation}
\sigma_i = \sqrt{\Sigma_{ii}}
\label{eq:one_sigma_err}
\end{equation}
The off-diagonal elements of the covariance matrix contain information about the correlation between different parameters. We compute the Fisher matrix in the following 9-dimensional parameter space 
\begin{equation}
    \bm{\theta} \equiv \{\ln \mathcal{M}, \eta, \ln d_L, \cos \iota, t_c, \phi_c, \alpha, 
    \sin \delta, \psi \}
\end{equation}
Equation \eqref{eq:one_sigma_err} assumes that the likelihood is not cut by the prior boundaries, which can happen in real situations. In order to mimic this, we draw a large number of random samples from the 9-dimensional Gaussian likelihood (computed using the the Fisher matrix; see Eq. \eqref{eq:err_prob}), and discard those samples that lie outside the prior boundaries. From the remaining samples, we compute the marginalized 1-dimensional posteriors in $d_L$ as well as $\iota$, and estimate the width of the 90\% confidence regions centered around their median values. These are considered as our error estimates in $d_L$ and $\iota$. We use the priors for the parameters as shown in Table \ref{tab:priors}.

 \begin{table}
 \begin{center}
\caption{\small{The priors on parameters for which the Fisher matrix is computed. Here $\mathrm{U(a, b)}$ denotes the uniform probability between $a$ and $b$}. The luminosity distance, mass, time, and all the angles are measured in Mpc, $M_{\odot}$, seconds, and radians respectively.}
\label{tab:priors}
\begin{tabular}{|c|c|}
\hline
\textbf{Parameter ($\theta$)} & \textbf{Prior}       \\ \hline
$\ln (d_L/\mathrm{Mpc}) $                     & $\mathrm{U}(0, 11.5)$ \\ \hline
$\cos \iota$                  & $\mathrm{U}(-1, 1)$           \\ \hline
$t_\mathrm{c}$ (sec)                     & $\mathrm{U}(-1, 1)$         \\ \hline
$\phi_\mathrm{c}$                      & $\mathrm{U}(0, 2\pi)$         \\ \hline
$\ln (\mathcal{M}/M_\odot)$            & $\mathrm{U}(0, 4.6)$ \\ \hline
$\eta$                        & $\mathrm{U}(0, 0.25)$         \\ \hline
$\sin \delta$                 & $\mathrm{U}(-1, 1)$           \\ \hline
$\alpha$                      & $\mathrm{U}(0, 2\pi)$         \\ \hline
$\psi$                        & $\mathrm{U}(0, 2\pi)$         \\ \hline
\end{tabular}
\end{center}
\end{table}

Computing the $90\%$ confidence regions in multi-dimensional parameter space (e.g., the three dimensional volume in sky) is a bit more computationally complex. Hence we resort to the following approximation: From the posterior samples that generate as described above, we compute the covariance matrix $\bar{\Sigma}^\mathrm{3D}$ in three dimensions ($\alpha, \sin \delta$, and $d_L$) numerically~\footnote{If the original 9-dimensional likelihood is not cut by the prior boundaries, the marginalized posterior in 2 dimensions will also be a Gaussian, which is fully described by this covariance matrix. However, the prior boundaries can cut the 9-dimensional Gaussian. We, still approximate the 3D distributions to be Gaussians, described by the covariance matrix $\bar{\Sigma}^\mathrm{3D}$.}. Then, the errors in the 3-dimensional sky-volume $\Delta V$ are given in terms of this marginalized covariance matrix $\bar{\Sigma}^\mathrm{3D}$ as

 
\begin{equation}
\Delta V = \frac{4}{3} \pi
\sqrt{
\renewcommand\arraystretch{2} 
\begin{vmatrix}
\bar{\Sigma}^\mathrm{3D}_{d_L d_L} &  d_L \bar{\Sigma}^\mathrm{3D}_{d_L \alpha} &  d_L \bar{\Sigma}^\mathrm{3D}_{d_L \sin\delta} \\
d_L \bar{\Sigma}^\mathrm{3D}_{d_L \alpha} &  d_L^2 \bar{\Sigma}^\mathrm{3D}_{\alpha \alpha} & d_L^2 \bar{\Sigma}^\mathrm{3D}_{\alpha \sin\delta} \\
d_L \bar{\Sigma}^\mathrm{3D}_{d_L \sin\delta} &  d_L^2 \bar{\Sigma}^\mathrm{3D}_{\alpha \sin\delta} & d_L^2 \bar{\Sigma}^\mathrm{3D}_{\sin\delta \sin\delta}
\end{vmatrix}}
\label{eq:one_sigma_vol}
\end{equation}
Note that equation \eqref{eq:one_sigma_vol} provides the uncertainties in the localization volume at $68\%$ confidence. To convert these 1-$\sigma$ errors into errors at confidence level $C$, we will have to multiply them by a scaling factor ($\beta$) \citep{credible_interval}. We have estimated this scaling factor in different dimensions in appendix \ref{appendix:conf_interval}. All the error estimates in this paper correspond to the $90\%$ credible interval unless otherwise stated. The above errors at $90\%$ confidence level will be,
\begin{eqnarray}
    \Delta V_{90\%} &=& (\beta_3)^3 \Delta V
\end{eqnarray}
where $\beta_n$ corresponds to the scaling factor at $90\%$ confidence level in $n$-dimensions. We are more interested in finding the number of galaxies localized ($\Delta N$) that can be the potential host of the merger. Therefore
\begin{eqnarray}
    \Delta N_{90\%} = n_{\mathrm{galaxy}} \ \Delta V_{90\%}
\end{eqnarray}
where $n_{\mathrm{galaxy}}$ is the number density of the galaxies in the universe.

The inversion of the Fisher matrix (for computing the covariance matrix) is performed numerically using the LU decomposition method in $\mathsf{mpmath}$ library with arbitrary precision \citep{mpmath} in $\mathsf{Python}$. Since the numerical techniques used in inverting $\Gamma$ may affect the inversion accuracy, we have to define a fiducial threshold of inaccuracies above which the results can not be trusted. This can be checked by inferring the deviation of the identity matrix from the multiplication of the inversion of the covariance matrix to the original Fisher matrix. The measure of accuracy can be defined as $\epsilon_{\mathrm{inv}} = \mathrm{max}_{i,j} |\Gamma_{ik} \Sigma_{kj} - \delta_{ij}|$ \citep{inversion_accuracy}. We find the values of $\epsilon_{\mathrm{inv}} \lesssim 10^{-8}$ in our calculations which is well within the acceptable limits (see \cite{inversion_accuracy}).

\subsection{Detector networks}
\begin{figure}
  \begin{center}
    \includegraphics[width=1\columnwidth]{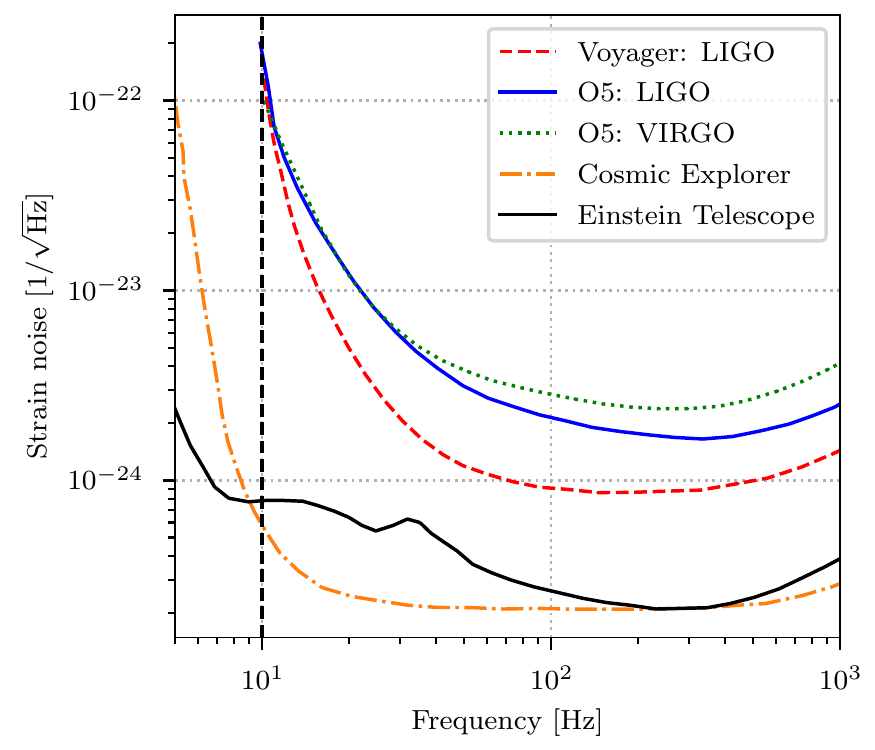} 
  \end{center}
  \caption{Sensitivities of the detectors in various observing scenarios. We have assumed that KAGRA detector will be operating at the same sensitivity as Virgo detector in the O5.}
  \label{fig:psd}
\end{figure}

In this work, we mainly focus on three observing scenarios: (i) O5: 5 detector network consisting of three LIGO detectors (LIGO-Hanford, LIGO-Livingston, and LIGO-India), one Virgo detector, and one KAGRA detector with sensitivities projected to the fifth observing run \citep{observer_summary, LIGO-INDIA, SaleemLIGOIndia}. (ii) Voyager: network of the same five detectors as in O5 but all three LIGO detectors upgraded to Voyager sensitivities \citep{Adhikari_voy, Voyager_PSD} while Virgo and KAGRA working at their O5 sensitivities. (iii) 3G: network of three detectors, one Einstein telescope \citep{ET} and two Cosmic Explorers \citep{CE}, with sensitivities projected for the third-generation detectors. In O5 and Voyager scenario, the detectors are sensitive above a low-frequency cutoff of $10$Hz below which, the sensitivity degrades rapidly due to seismic noise. This low frequency wall is pushed down to 5Hz in the case of 3G detectors. Fig. \ref{fig:psd} shows the amplitude spectral densities of the detectors in various observing scenarios. Note that the Einstein telescope has been assumed to be of L-shape in our analysis.


\section{Results}\label{sec:results}
	We have used a non-spinning multipolar waveform model by \cite{Mehta_HMs} which is calibrated against the numerical relativity simulations. We use two subdominant multipoles $\ell=3,m=\pm 3$ and $\ell=4,m=\pm 4$ in addition to the dominant mode $\ell=2,m=\pm 2$ throughout the analysis. The derivatives of the waveform with respect to the binary source parameters $\vec{\theta}$, which are being used to compute Fisher information matrix, have been calculated in $\mathsf{Mathematica}$ \citep{Mathematica_c} analytically to avoid any numerical errors due to finite differencing. We compute the expected uncertainties in the luminosity distance and orbital inclination as a function of the component masses (after fixing the sky location and polarization angle). We also simulate a population of binaries (with fixed masses, but different sky location and polarization angles) and compute the distribution of expected uncertainties. 

\subsection{Expected uncertainties as a function of component masses}

\begin{figure*}
  \centering
    \includegraphics[width=1.0\textwidth]{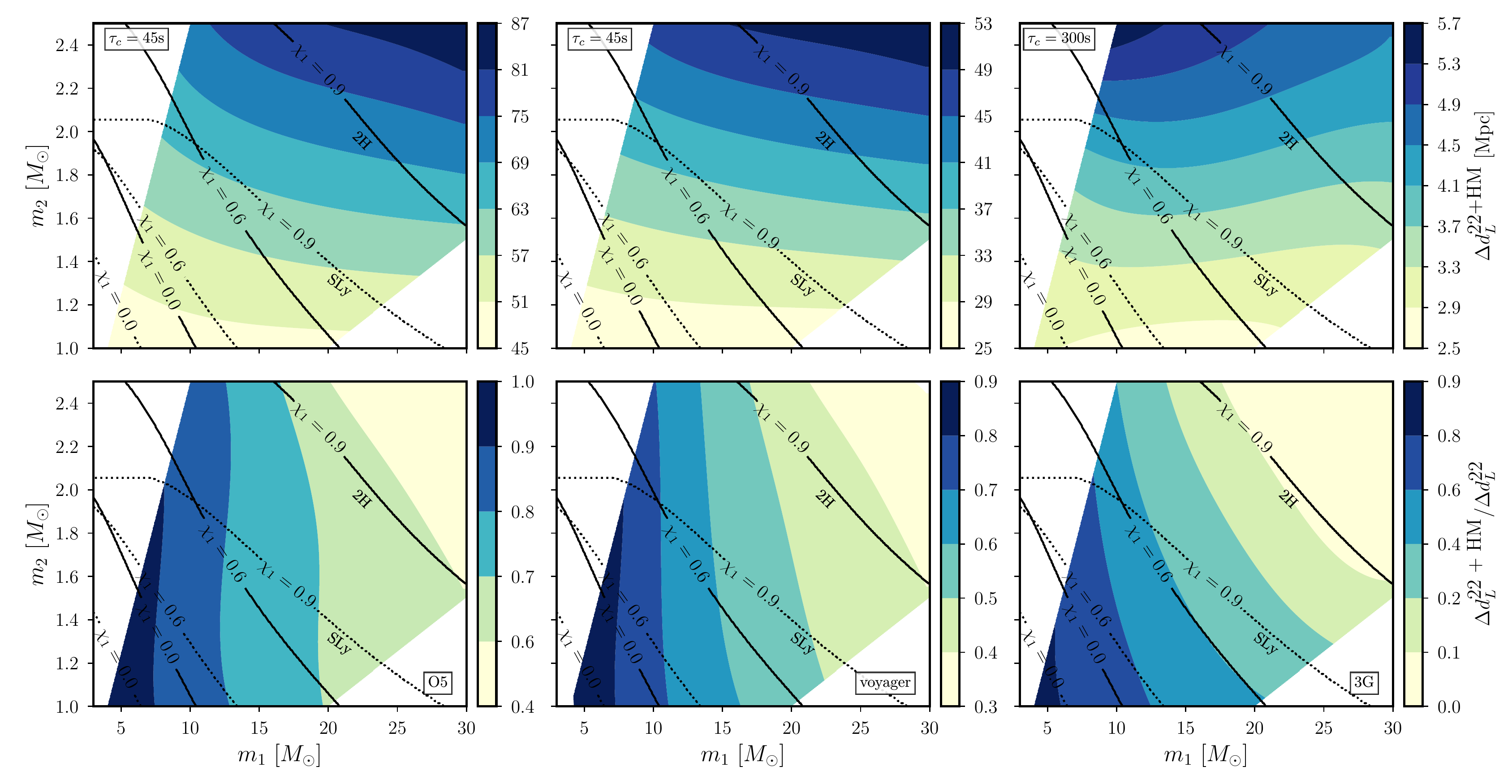}
    \caption{\textit{Top panel:} Expected uncertainties in estimating the luminosity distance $d_L$ (at $90\%$ confidence) with the inclusion of higher modes for non-spinning compact binary mergers, located at $100$Mpc, in three observing scenarios. We consider the component masses: $m_1 = 4-30 M_{\odot}, m_2 = 1-2.5 M_{\odot}$.  Other parameters of the binary systems are assumed to be fixed at their ``optimal'' values (values producing the best estimates of the distance) including $\iota = 60$deg. The black contours demarcate the regions of binaries, that will have potential EM emission i.e. non-zero ejecta mass in the merger, for various values of the spin ($\chi_1$) of the primary \citep{FoucartEMB1}. The black solid and dotted line contours correspond to the equations of state, 2H and SLy for the neutron star.  \textit{Bottom panel:} fractional improvements in the luminosity distance measurements with the inclusion of higher modes, relative to the measurements  carried out using only dominant mode. In O5 (voyager)[3G] scenario, the luminosity distance measurements improve by a factor of $\sim 1-1.5 (1.1-2)[1.1-5]$, $45 (45) [300]$ seconds prior to the merger.}
    \label{fig:d_errs_density_mass_var}
\end{figure*}

\begin{figure*}
\centering
	\includegraphics[width=1.0\textwidth]{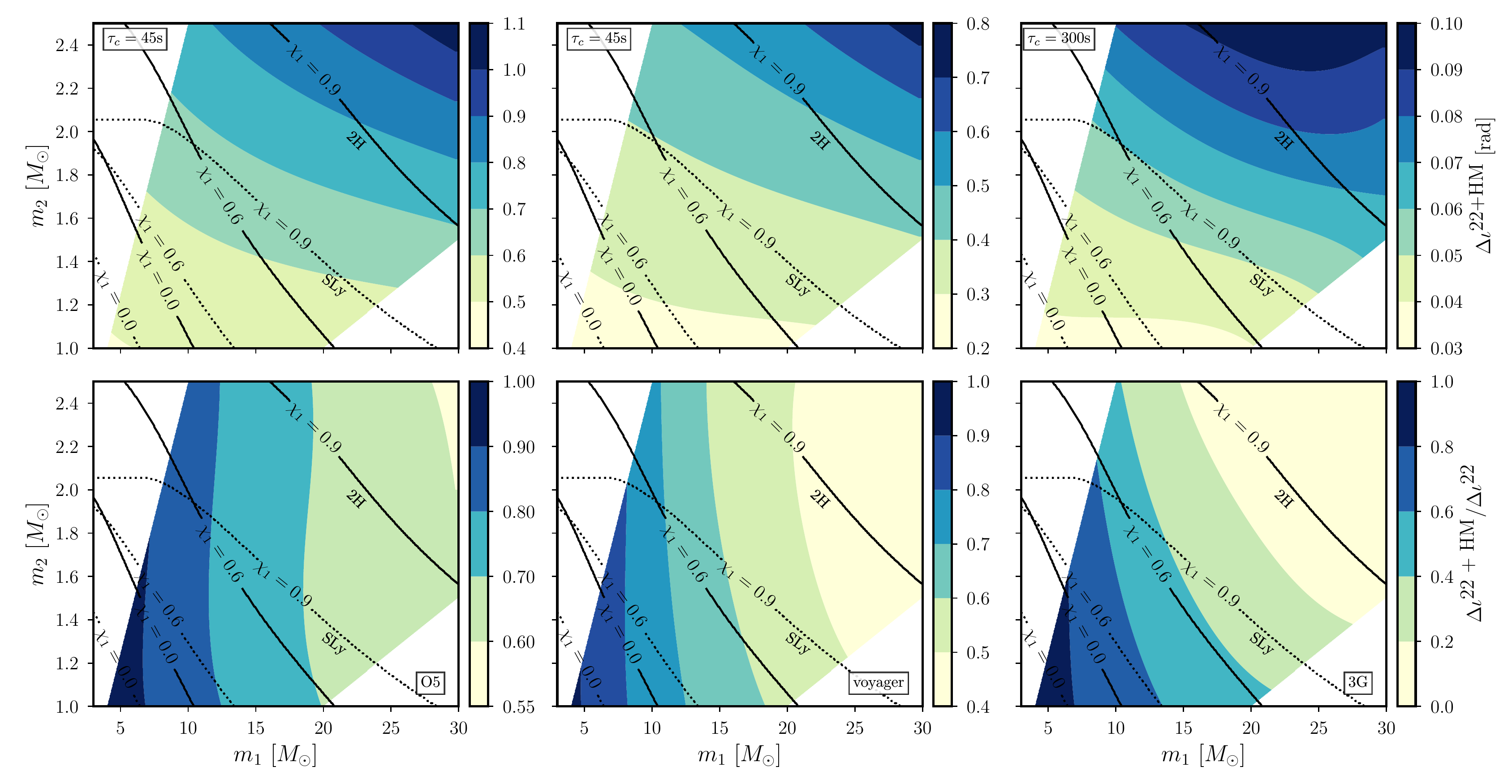} 
    \caption{Same as in plot (\ref{fig:d_errs_density_mass_var}) but we show improvements in the measurement of orbital inclination ($\iota$) of compact binary mergers with the inclusion of higher modes. The orbital inclination measurements improve by a factor $\sim 1-1.5(1-2)[1-3]$, with the inclusion of higher modes, for an early warning time of $45(45)[300]$ seconds, in O5(Voyager)[3G].}
    \label{fig:incl_errs_density_mass_var}
\end{figure*}

\begin{figure*}
  \centering
  \includegraphics[width=1.0\textwidth]{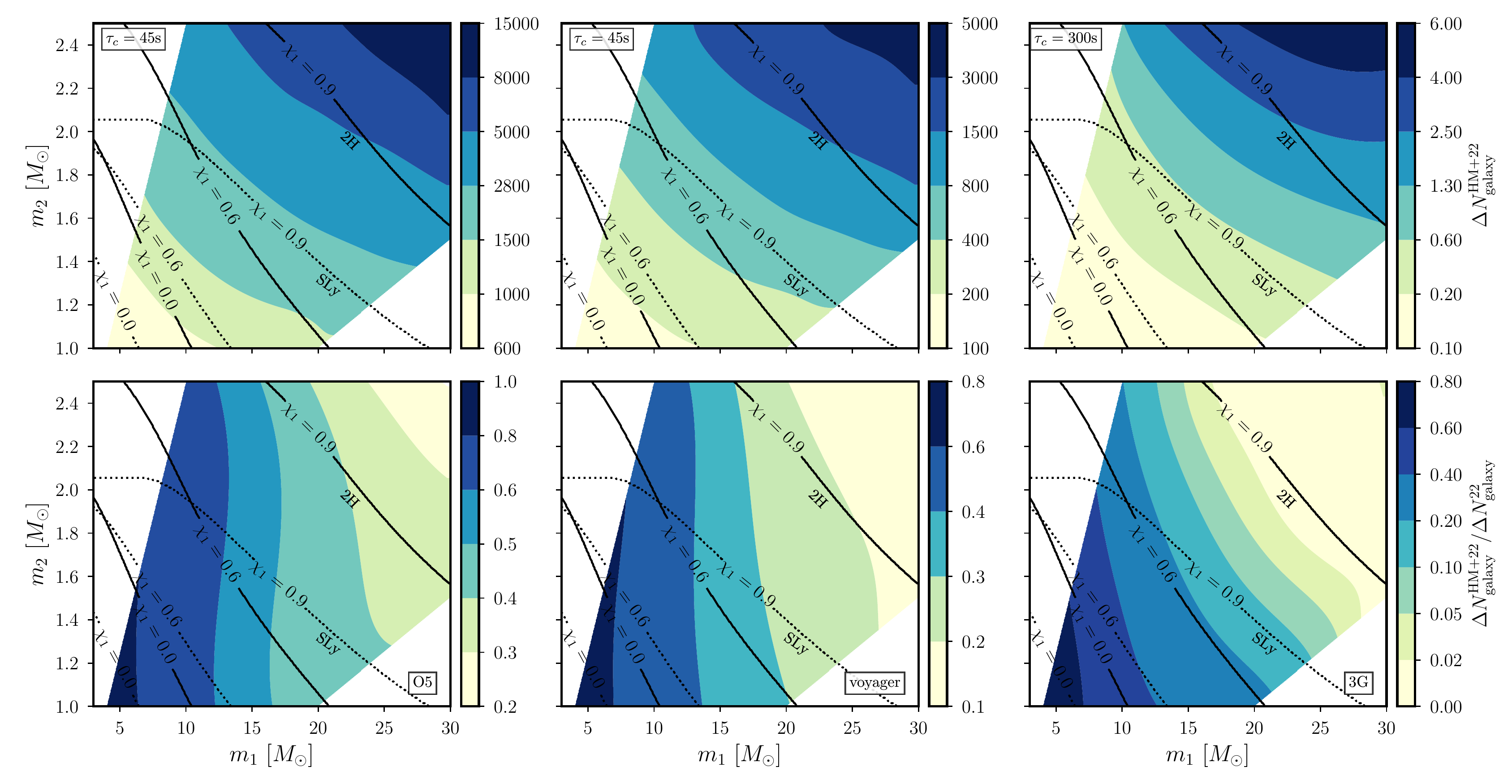}
  \caption{\textit{Top panel:} the total number of galaxies localized (at $90\%$ confidence) that can be potential hosts of the compact binary merger with the inclusion of higher modes. This number can be as small as $\sim 1000(200)[0.2]$ at $45(45)[300]$ seconds before the merger for binary systems that will have potential EM counterpart in O5 (Voyager) [3G]. \textit{Bottom panel:} the reduction in the number of potential host galaxies with the inclusion of higher modes relative to only dominant mode. This reduction factor can be as $\sim 1-2.5(1.2-4)[1.2-10]$ for an early warning time of $45(45)[300]$ seconds before the merger in O5(Voyager)[3G].}
  \label{fig:density_num_galaxies}
\end{figure*}

The first set of results are generated by simulating signals in a grid of masses spanning $m_1=4-30 M_{\odot}, m_2=1-3 M_{\odot}$ which corresponds to the mass range of NSBH binaries. We have chosen other extrinsic parameters as follows: inclination angle $\iota=60$ deg., luminosity distance $d_L = 100 \text{Mpc}$, and sky-location ($\alpha, \delta$) and polarization angle ($\psi$) corresponding to the values which provide most precise estimates of localization skyarea of the source. 

In Figure \ref{fig:d_errs_density_mass_var}, we show the improvements in the measurement of luminosity distance ($d_L$) with the inclusion of higher modes at different fiducial early warning times in various observing scenarios. We quantify the improvement in terms of the ratio of the width of the $90\%$ credible interval in the marginalized posteriors of $d_L$. We find that in O5(Voyager), the measurement of luminosity distance is improved by a factor as large as $\sim 1-1.5 (1-2)$ with the inclusion of higher order modes, 45 seconds before the merger. A significant number of these mergers are also expected to produce an EM counterpart \footnote{Throughout this paper, we have assumed that a merger producing a non-zero remnant mass outside the innermost stable circular orbit (ISCO) of the final black hole will be EM bright \citep{FoucartEMB1}}. The possibility of an EM emission in a merger increases with the spin of the primary component (black hole in case of a NSBH) in a binary. This is a consequence of decreasing ISCO radius with increasing spins, and hence leading to higher chances of tidal disruption of matter happening outside the ISCO. The EM bright nature of the compact binary merger also depends on the EoS of the neutron star component. The stiffer the EoS of a neutron star, the greater are the chances of the binary being EM-bright.

In 3G, the luminosity distance measurements can improve by a factor up to $\sim 1.1 - 5$ at 300 seconds before the merger for a significant number of binaries. The early warning time for 3G observing scenario has been chosen as 300 seconds to keep in mind the fact that 3G detectors will be sensitive to frequencies as low as 5Hz (see Figure \ref{fig:psd}). This will give us a longer early-warning time. Figure \ref{fig:incl_errs_density_mass_var} shows the orbital inclination angle ($\iota$) measurements at the same early warning times for all three observing scenarios. The improvement factors in the measurement of $\iota$ are $\sim 1-1.5(1-2)[1-3]$ in O5(Voyager)[3G] for many compact binary mergers with significant fraction of them being EM bright similar to that of luminosity distance measurements. Although the improvements in orbital inclination measurements are not that significant, it can help in better constraining the orientation of a beamed EM emission, if there exists any, from the NSBH mergers more accurately than what could have been done using only the dominant mode.

 Given the measurements of luminosity distance and localization skyarea, we can also estimate the error volume in which a source can be localized. This error volume further can be translated into the number of galaxies given the number density of galaxies in the universe. An estimate of the number of galaxies in this error volume is one of the most important parameters in which the astronomers will be interested while searching for EM counterparts of a binary merger.

 In Figure \ref{fig:density_num_galaxies}, we show the expected number of galaxies ($\Delta N$) localized, at $90\%$ confidence, with the inclusion of higher modes and also the improvements relative to only dominant mode estimates, at different EW times in various observing scenarios. We have assumed the number density of galaxies in the universe as $n_{\mathrm{galaxy}} = 0.01 \mathrm{Mpc}^{-3}$ \citep{galaxy_density}. In the best case scenario, the number of galaxies localized ($\Delta N$), with the inclusion of higher modes, can be as small as $\sim 1000(200)[0.2]$ at $45(45)[300]$ seconds before the merger in O5(Voyager)[3G]. Including higher modes apart from the dominant mode can help in reducing the number of galaxies localized by a factor as large as $\sim 1-2.5(1.2-4)[1.2-10]$ for an early warning time of $45(45)[300]$ seconds before the merger in O5(Voyager)[3G]. 3G detectors will almost always be able to pin point the source to a single galaxy for a significant number of binary mergers that will also potentially be EM bright.

\subsection{Variation of sky location and polarization angle}

\begin{figure*}
  \centering
  \subfloat[Distributions of the measurement errors in $d_L$ and $\iota$ for 1000 compact binary mergers with varying sky-locations and polarizations. The \textit{left column} shows the cumulative distributions of $d_L$ errors. Solid (dashed) lines correspond to errors computed including (neglecting) the higher mode contributions for three different early warning times of $20, 30,$ and $45$ seconds. In O5 (Voyager) scenario, the median $d_L-$errors are $44-81 (25-51)$ Mpc at $20-45$ seconds before the merger respectively. The \textit{right column} shows the  reductions of $d_L-$errors with the inclusion of higher modes relative to only dominant mode measurements. For over $50\%$ of the binaries, inclusion of higher modes will cause the $d_L$ errors to reduce by a factor of $\sim 1.2-2 (1.3-5)$ in O5 (Voyager).]{
    \includegraphics[width=0.9\textwidth]{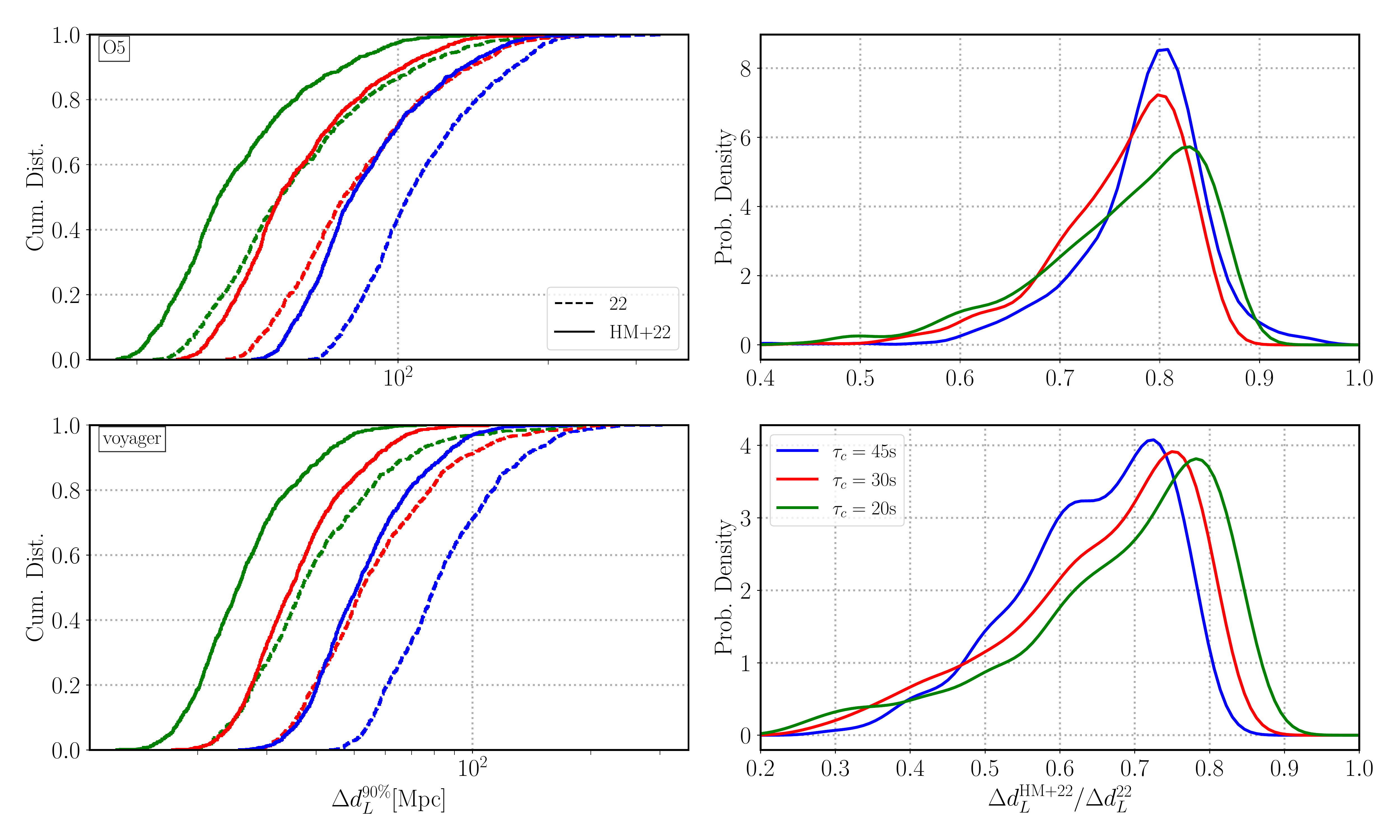}
    \label{fig:d_errs_sky_loc_var}}
    \\
  \subfloat[Same as figure \ref{fig:d_errs_sky_loc_var} but histograms show the inclination errors here. For over $50\%$ of the binaries, the errors are reduced by a factor of $\sim 1.2-2 (1.3-5)$ in O5 (Voyager).]{
    \includegraphics[width=0.9\textwidth]{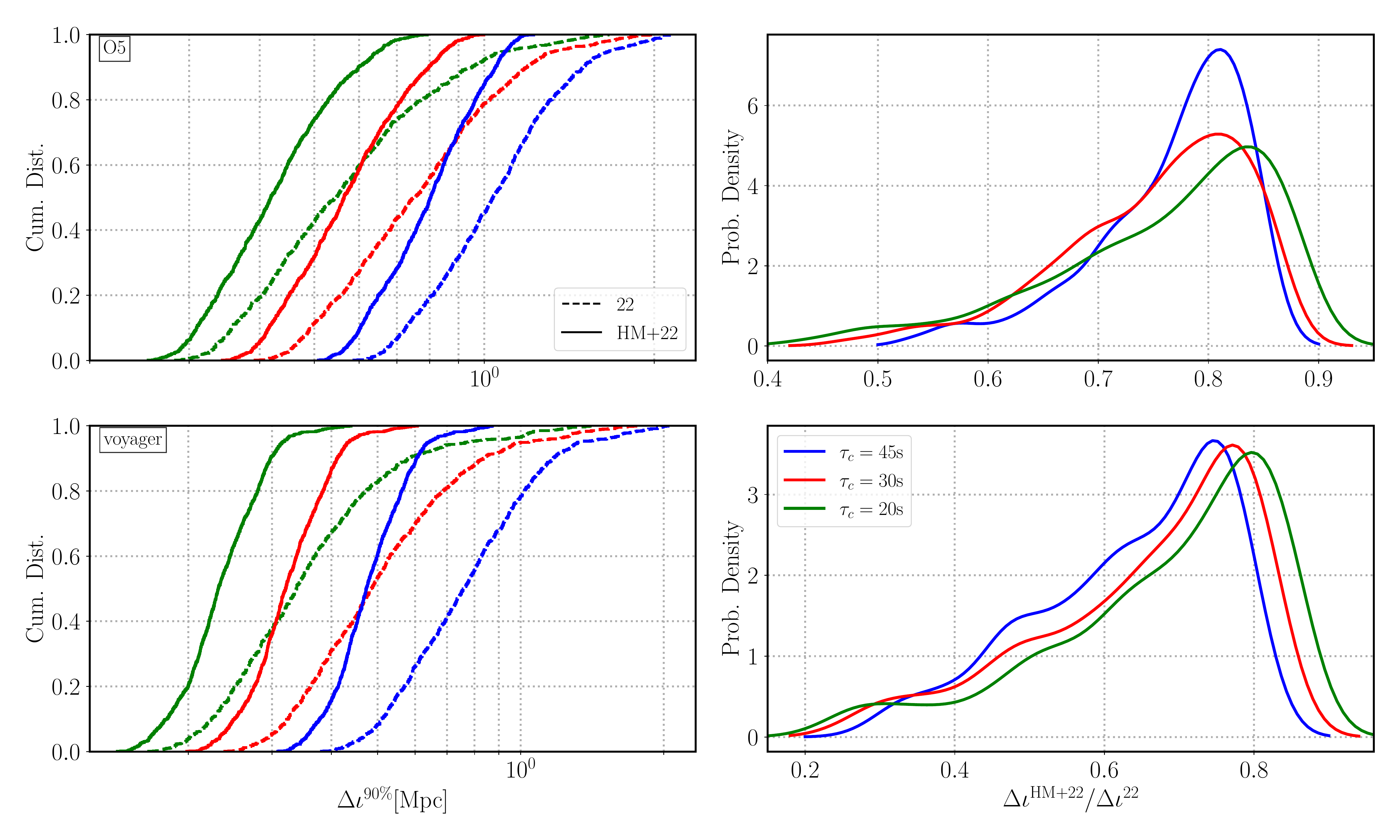} 
    \label{fig:incl_errs_sky_loc_var}}
\end{figure*}

In the final set of results, we look at the variation of the sky-location and polarization on the error measurements of luminosity distance and orbital inclination while keeping the rest of the parameters fixed. The fixed parameters are again $m_1 = 15 M_{\odot}, m_2 = 1.5 M_{\odot}, \iota= 60$ deg., and $d_L = 100$Mpc. We simulate 1000 uniformly distributed sky-locations and polarizations of compact binary mergers. The left column in the Figure \ref{fig:d_errs_sky_loc_var} shows the cumulative histograms of luminosity distance measurement errors with and without the inclusion of higher modes at three different early-warning times: 45, 30, and 20 seconds in both O5 and Voyager observing scenarios. The right column corresponds to the fractional improvements in the measurement errors of luminosity distance with the inclusion of higher modes. Median values of these distributions are tabulated in Table \ref{tab:sky_loc_var_tab}.  The improvements are better at early times than near the merger since the relative contribution of higher modes is larger at earlier times\footnote{This is because the higher modes oscillate in the most sensitive frequency band of the detector, while the dominant mode is largely buried in the low-frequency noise.}. Similarly, Figure \ref{fig:incl_errs_sky_loc_var} shows the improvements in the measurement of inclination angle while varying the sky-locations and polarizations. 


\begin{table}
\caption{\small{Median values of the distribution of uncertainties in the estimation of the luminosity distance $d_L$ and inclination angle $\iota$ using templates including the contribution of higher modes. These correspond to two observing scenarios (O5 and Voyager) and three different early warning times (20, 30, and 45 seconds). The median improvement factors (as compared to the estimates using only 22 mode) are shown in parentheses.}} 
\label{tab:sky_loc_var_tab}
\begin{tabular}{|c|ccc|ccc|}
\hline
\multicolumn{1}{|l|}{}                                                                                   & \multicolumn{3}{c|}{\textbf{O5}}                                                                                                                                                                                & \multicolumn{3}{c|}{\textbf{Voyager}}                                                                                                                                                                           \\ \hline
$\tau_c${[}seconds{]}                                                                                    & \multicolumn{1}{c|}{20}                                                    & \multicolumn{1}{c|}{30}                                                    & 45                                                    & \multicolumn{1}{c|}{20}                                                    & \multicolumn{1}{c|}{30}                                                    & 45                                                    \\ \hline
\begin{tabular}[c]{@{}c@{}}$ \Delta d_L^{\mathrm{HM+22}}$ in Mpc\\  (Reduction factor)\end{tabular}      & \multicolumn{1}{c|}{\begin{tabular}[c]{@{}c@{}}44\\  (1.2)\end{tabular}}     & \multicolumn{1}{c|}{\begin{tabular}[c]{@{}c@{}}58 \\ (1.3)\end{tabular}}     & \begin{tabular}[c]{@{}c@{}}81\\  (1.2)\end{tabular}    & \multicolumn{1}{c|}{\begin{tabular}[c]{@{}c@{}}25 \\ (1.4)\end{tabular}}   & \multicolumn{1}{c|}{\begin{tabular}[c]{@{}c@{}}35\\  (1.5)\end{tabular}}   & \begin{tabular}[c]{@{}c@{}}51\\  (1.5)\end{tabular}    \\ \hline
\begin{tabular}[c]{@{}c@{}}$\Delta \iota^{\mathrm{HM+22}}$ in radians\\  (Reduction factor)\end{tabular} & \multicolumn{1}{c|}{\begin{tabular}[c]{@{}c@{}}0.41 \\ (1.3)\end{tabular}} & \multicolumn{1}{c|}{\begin{tabular}[c]{@{}c@{}}0.55 \\ (1.3)\end{tabular}} & \begin{tabular}[c]{@{}c@{}}0.77\\  (1.3)\end{tabular} & \multicolumn{1}{c|}{\begin{tabular}[c]{@{}c@{}}0.23\\  (1.4)\end{tabular}} & \multicolumn{1}{c|}{\begin{tabular}[c]{@{}c@{}}0.31 \\ (1.4)\end{tabular}} & \begin{tabular}[c]{@{}c@{}}0.46\\  (1.5)\end{tabular} \\ \hline
\end{tabular}
\end{table}



\section{Summary and Outlook}\label{sec:conclusion}
	Joint GW-EM observations of EM-bright CBCs promise to shed light on the complex physics of the merger and associated phenomena.  Among them, BNS mergers within $\mathcal{O}(100)$ Mpc are the most likely to produce observable EM counterparts. 
It is not surprising that a number of GW early-warning efforts targeted at BNS mergers are currently under way \citep{SachdevEW, MageeEW, MageeEW2, Nitz_EW}. 

On the other hand, EW studies focussed on potentially EM-Bright NSBH binaries have only recently gained attention \citep{nsbh_EW2, nsbh_EW1}, due primarily to the fact that their inspiral duration within the frequency band of ground-based detectors is significantly shorter than BNSs.  Nevertheless, our previous work \citep{Kapadia2020, Singh_EW2021} demonstrated that including higher harmonics of GW radiation in templated low-latency searches could considerably increase the duration of the signal in-band. As a result, we showed that EW detection and sky-localisation could be improved considerably in future observing runs (O5, Voyager, 3G).  

We follow-up our previous work by demonstrating the EW benefits of including higher modes in reducing the localisation sky-volume, while also improving estimates of orbital inclination. Adopting the Fisher Matrix analysis, we find that for a range of potentially EM-Bright NSBH systems located at 100 Mpc, the error-bars on the distance reduce by a factor of $\sim 1-1.5 (1.1-2)[1.1-5]$ at early warning times of 45 (45) [300] seconds, pertaining to observing runs O5 (Voyager) [3G].  We then pick a fiducial NSBH binary with masses $15-1.5 M_{\odot}$ which could potentially be EM-Bright for a moderately spinning BH, and vary its sky-location and polarization angle. Of the $1000$ randomly selected locations and polarizations, we find that the median $d_L-$errors range from $44-81 (25-51)$ Mpc at $20-45$ seconds before the merger. These correspond to $d_L-$error reduction factors of  $\sim 1.2 (1.5)$ upon the inclusion of higher modes, in O5 (Voyager) scenario. 

Improved early-warning estimates of the luminosity distance and sky-volume could aid astronomers in determining their follow-up strategy. Different EM telescopes have limiting distances to which they can probe. Of course, increasing the exposure time would enable them to probe larger distances. However, for transient GW events that are to be followed up in early-warning time, large exposure times are not feasible.  

Furthermore, even if a telescope had a sufficient depth of view, it would need to slew to the appropriate sky-location, scan the localization volume, and point at the NSBH before it merges.  Assisted by a galaxy catalog, as well as a coordinated search involving multiple telescopes, capturing the EM-counterpart at its onset could in principle be achieved \citep{SlewingStrategy, Grandma}.  With this in mind, we also highlight that using higher modes, the number of galaxies to be searched over can be reduced by as much $\sim 1-2.5(1.2-4)$, 45 seconds before merger for NSBH systems located at $100$ Mpc in the O5(Voyager) scenario. 

It would be worth mentioning here that the Fisher Matrix likelihood, being an expansion truncated at quadratic order of the full GW parameter estimation likelihood, can deviate from the full likelihood non-trivially. While this is true in general for lower SNR events, we also show in Appendix \ref{appendix:analytical_FM_est}, that this could occur at small inclinations as well. For the fiducial inclination angle of $60$ degrees chosen in this work (see Figures \ref{fig:Bay_Fish_l_22}), we find that the full and Fisher Matrix likelihoods agree well.

\section*{Acknowledgements}
We thank Geoffrey Mo for carefully reviewing our manuscript and providing useful comments. MKS's, SJK's, MAS's and PA's research was supported by the Department of Atomic Energy, Government of India, under project no. RTI4001. In addition, SJK’s work was
supported by a grant from the Simons Foundation
(Grant No. 677895, R. G.). PA's research was funded by the Max Planck Society through a Max Planck Partner Group at ICTS-TIFR and by the Canadian Institute for Advanced Research through the CIFAR Azrieli Global Scholars program. Numerical calculations reported in this paper were performed using the Alice cluster at ICTS-TIFR, and ``powehi" workstation at the Department of Physics, IIT Madras.

\bibliography{references}
\newpage

\onecolumn
\section*{}\label{sec:appendix}
	\begin{appendix}
\section{Analytical Fisher matrix errors estimation}
\label{appendix:analytical_FM_est}

\begin{figure*}
  \includegraphics[trim=0 10 10 15, clip, width=0.48\textwidth]{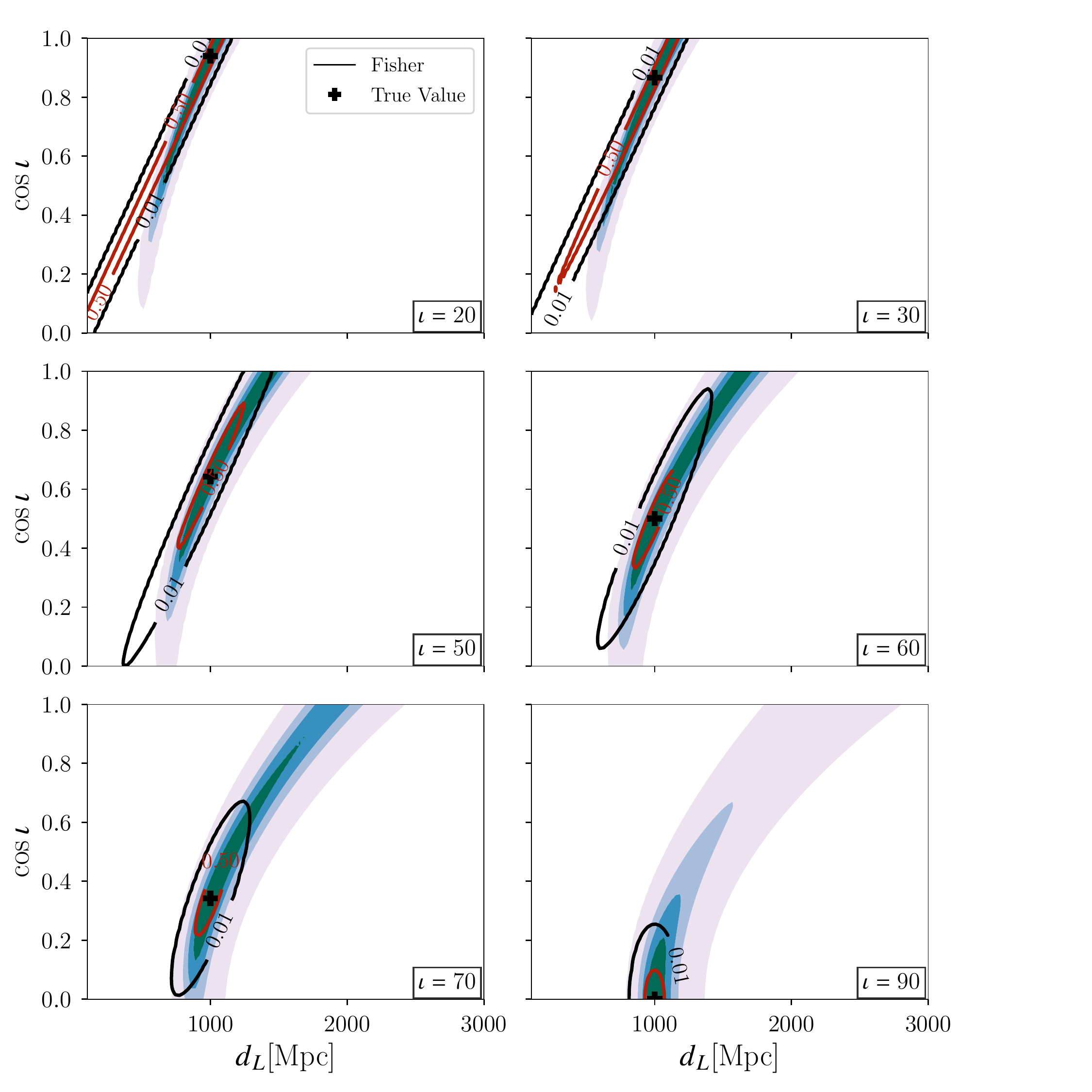}
  \includegraphics[trim=0 10 10 15, clip, width=0.48\textwidth]{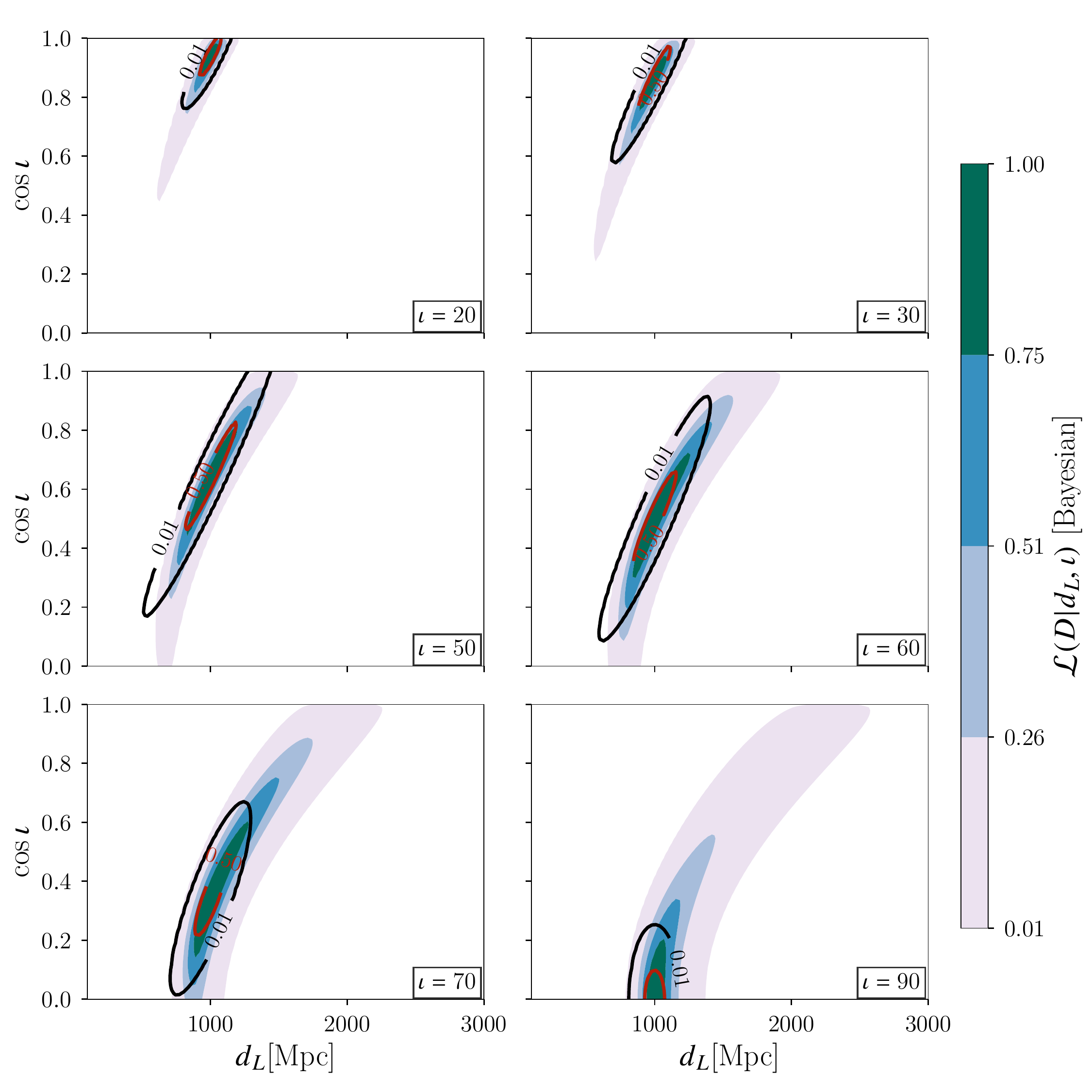}
  \caption{Fisher and Bayesian likelihood comparison in distance and inclination angle plane while varying inclination angle with only dominant mode (left) and multipolar (right) waveform.}
  \label{fig:Bay_Fish_l_22}
\end{figure*}

As discused in section \ref{sec:method}, the Fisher matrix provides a quadratic (Gaussian) approximation of the true Bayesian likelihood. While this might work for most cases, this might not provide a good approximation when the posteriors ore multimodal, or when they have non-trivial shapes. We found that the Gaussian likelihood in $\cos \iota-d_L$ provided by the Fisher matrix is not a good approximation of the true Bayesian likelihood for certain choices for $\iota$ (Fig \ref{fig:Bay_Fish_l_22}). This is especially the case for the analysis using only dominant mode when the true likelihood is significantly wide, due to correlations between parameters. When the higher modes are included, this reduces the correlation between $\iota-d_L$ and reduces the size of the likeliliood, rendering the Gaussian approximation more accurate. We restrict our study to the values of $\iota$ where the approximation is a reasonable one ($\iota = 60$). To get an understanding of the deviation of Fisher matrix likelihood from the true likelihood at low inclination angles, we perform some analytical calculations for the Fisher matrix analysis with dominant mode.


For the dominant mode of GW radiation, the $+$ and $\times$ polarizations of a GW signal in frequency domain \citep{Mehta_HMs} can be written as
\begin{eqnarray}
\tilde{h}_{+}(f) &=&  \frac{1 + \cos^2 \iota}{2 d_L} \tilde{h}_{22} (f; \vec{\lambda}), 
\label{eq:hplus} 
\\
\tilde{h}_{\times} (f) &=& \frac{-i}{d_L} (\cos \iota) \tilde{h}_{22} (f; \vec{\lambda}) \label{eq:hcross}
\end{eqnarray}
where $\tilde{h}_{22} (f; \vec{\lambda})$ depends only on the intrinsic parameters of the source. The total signal is
\begin{eqnarray}
\tilde{h}(f) = F_+ \tilde{h}_{+}(f) + F_{\times} \tilde{h}_{\times}(f),
\label{total_amp}
\end{eqnarray}
where $F_+ (\alpha, \delta, \psi)$ and $F_{\times}(\alpha, \delta, \psi)$ are antenna pattern functions of the detector. Considering only a single detector and substituting Eqs. (\ref{eq:hplus}) and (\ref{eq:hcross}) into Eq. (\ref{total_amp}), we get
\begin{eqnarray}
\tilde{h}(f) = \frac{e^{-\ln d_L}}{2} \left[ (1 + \cos^2 \iota) F_+ - i (2 \cos \iota ) F_{\times} \right] \tilde{h}_{22} (f; \vec{\lambda}).
\end{eqnarray}
Evaluating Fisher matrix (FM) requires the computation of derivatives of the waveform with respect to the parameters. Assuming the parameterization ($\ln d_L, \cos \iota$), we calculate the derivatives of $\tilde{h}(f)$ as
\begin{eqnarray}
\tilde{h}_{(\ln d_L)} (f)
 &=& \partial_{(\ln d_L)} \tilde{h}(f) = - \frac{e^{-\ln d_L}}{2} \left[ (1 + \cos^2 \iota) F_+ - i (2 \cos \iota ) F_{\times} \right] \tilde{h}_{22} (f; \vec{\lambda}) \\
\tilde{h}_{(\cos \iota)} (f) &=& \partial_{(\cos \iota)} \tilde{h}(f) =  \frac{e^{-\ln d_L}}{2} \left[ (2 \cos \iota) F_+ - 2 i F_{\times} \right] \tilde{h}_{22} (f; \vec{\lambda})
\end{eqnarray}
The FM elements for $\ln d_L$ and $\cos \iota$ are
\begin{eqnarray}
F_{(\ln d_L) (\ln d_L)} = \left( h_{(\ln d_L)}(f) \Big \rvert h_{(\ln d_L)}(f) \right) =  A \frac{e^{-2 \ln d_L}}{4} \left[ (1 + \cos^2 \iota)^2 F_+^2 + (2 \cos \iota )^2 F_{\times}^2 \right]
\end{eqnarray}
where $A = \left( \tilde{h}_{22} (f; \vec{\lambda}) \Big \rvert \tilde{h}_{22} (f; \vec{\lambda}) \right)$
 is constant throughout our estimates of FM elements since it depends only on intrinsic parameters which are fixed. Similarly, 
\begin{eqnarray}
F_{(\ln d_L) (\cos \iota)} 
= \left( h_{(\ln d_L)}(f) \Big \rvert h_{(\cos \iota)}(f) \right) = -A \frac{e^{-2 \ln d_L}}{4} (2 \cos \iota ) \left[ (1 + \cos^2 \iota) F_+^2 + 2 F_{\times}^2 \right] = F_{(\cos \iota)(\ln d_L)},
\end{eqnarray}
and
\begin{eqnarray}
F_{(\cos \iota) (\cos \iota)} = A \frac{e^{-2 \ln d_L}}{4}  \left[ (2 \cos \iota )^2 F_+^2 + (2 F_{\times})^2 \right]
\end{eqnarray}
The FM is given as
\begin{eqnarray}
F = \begin{pmatrix}
    F_{(\ln d_L) (\ln d_L)} & F_{(\ln d_L) (\cos \iota)} \\ \\
    F_{(\ln d_L) (\cos \iota)} & F_{(\cos \iota) (\cos \iota)}
    \end{pmatrix}.
\end{eqnarray}
We have to invert the above matrix to get the covariance matrix ($\Sigma$) which will render the errors and correlations between different parameters. Let us calculate the determinant of $F$, i.e. $|F|$ or $\mathrm{det} (F)$,  to find out if $F$ is invertible. 
\begin{eqnarray}
|F| = \mathrm{det} (F) = 4 F_+^2 F_{\times}^2 (1 - \cos^2 \iota)^2
\end{eqnarray}
Therefore, it is clear that $|F| \neq 0 $ in general (except for $\iota = 0$). The covariance matrix ($\Sigma$) is given as
\begin{eqnarray}
\Sigma 
 = \frac{A d_L^2}{F_+^2 F_{\times}^2 (1 - \cos^2 \iota)^2 }  \times \begin{pmatrix}
F_{\times}^2 +  (\cos^2 \iota) F_+^2  & 2 F_{\times}^2 + (1 + \cos^2 \iota) F_+^2   \\
\\
2 F_{\times}^2 + (1 + \cos^2 \iota) F_+^2 & 4 (\cos^2 \iota) F_{\times}^2 + (1 + \cos^2 \iota)^2 F_+^2
\end{pmatrix}
\end{eqnarray}
The term in the denominator $(1 - \cos^2 \iota)^2$ is a increasing function of $\iota \in [0, \pi/2]$. This is the dominating term which governs the overall increasing behaviour of all the covariance matrix elements (the errors and correlations) at low inclinations $\iota \lesssim 50$ deg. (see left plot in Fig. \ref{fig:Bay_Fish_l_22}). Focusing on errors in $\ln d_L$ and $\cos \iota$ the expressions are given by
\begin{eqnarray}
\sigma_{\ln d_L}  =  \frac{\sqrt{A} d_L}{F_+ F_{\times} (1 - \cos^2 \iota) } \sqrt{F_{\times}^2 +  (\cos^2 \iota) F_+^2}
\label{eq:dL_analy_err}
\end{eqnarray}
and
\begin{eqnarray}
\sigma_{\cos \iota}  =  \frac{\sqrt{A} d_L}{F_+ F_{\times} (1 - \cos^2 \iota) } \sqrt{4 (\cos^2 \iota) F_{\times}^2 + (1 + \cos^2 \iota)^2 F_+^2}
\label{eq:incl_analy_err}
\end{eqnarray}
which diverge at $\iota \sim 0$. Inclusion of higher modes will not lead to divergence factor $(1 - \cos^2 \iota)$ in the denominator of errors and correlations hence the Fisher matrix is a good approximation to the true likelihood even at low values of inclination angle (see right plot in Fig. \ref{fig:Bay_Fish_l_22}). 

\section{Confidence interval calculation}
\label{appendix:conf_interval}
A multivariate Gaussian distribution in N dimensions is given by
\begin{eqnarray}
p(\vec{x}) = \mathcal{N} \exp \left[ -\frac{1}{2}  (\vec{x}- \vec{\mu})^T \Sigma^{-1} (\vec{x} - \vec{\mu})\right]
\label{eq:multi_var_gau}
\end{eqnarray}
when positions $\vec{x} = \{x_1, x_2, ..., x_N\}$ and mean $\vec{\mu} = \{\mu_1, \mu_2, ..., \mu_N\}$ are given. When the covariance matrix $\Sigma$ is non-singular, it can be diagonalized to render the distribution as
\begin{eqnarray}
p(\vec{x}) = \mathcal{N} \exp \left[ - \sum_{i=1}^{N}\frac{(x_i - \mu_i)^2}{2 \sigma_i^2}  \right]
\end{eqnarray}
where $\sigma_i^2$'s are the eigenvalues of the covariance matrix $\Sigma$. The aim is to find the $C\%$ confidence region of this multivariate distribution. Let us look at this problem in 3-dimensions. The probability density is
\begin{eqnarray}
p(x,y,z) = \frac{1}{(2 \pi)^{3/2} \sigma_x \sigma_y \sigma_z} \exp\left[ - \frac{(x - \mu_x)^2}{2 \sigma_x^2} - \frac{(y - \mu_y)^2}{2 \sigma_y^2} - \frac{(z - \mu_z)^2}{2 \sigma_z^2} \right].
\label{eq:diag_prob_3d}
\end{eqnarray}
To estimate the volume of the ellipsoid at a particular confidence interval, we have to perform the integration on ellipsoidal symmetry. Just to estimate the scaling factor $\beta_3$ of different principle axes, let us choose the parametrization as follows:
\begin{eqnarray}
x & = & \mu_x + \sigma_x r \sin \theta \cos \phi \\
y & = &\mu_y + \sigma_y r \sin \theta \sin \phi \\
z & = &\mu_z + \sigma_z r  \cos \theta
\end{eqnarray}
where $r>0$ parametrizes the concentric ellipsoids, $\theta$ and $\phi$ are spherical polar angles. The volume element in these coordinates can be written as $dV =  \sigma_x \sigma_y \sigma_z r^2 \sin \theta dr d\theta d\phi$. In these coordinates, Eq. (\ref{eq:diag_prob_3d}) reduces to
\begin{eqnarray}
p(r, \theta, \phi) = \frac{1}{(2 \pi)^{3/2}} e^{-r^2/2}
\label{eq:prob_3d_sph_symm}
\end{eqnarray}
Let us assume that sphere of radius $\beta_3$ encloses a probability $C$, then
\begin{eqnarray}
C = \frac{1}{(2 \pi)^{3/2}} \int_0^{\beta_3} r^2 dr \int_0^{\pi} \sin \theta \ d\theta \int_0^{2 \pi} d\phi \ e^{-r^2/2}   
\label{eq:sph_confidence}
\end{eqnarray}
Or
\begin{eqnarray}
C = \sqrt{\frac{2}{\pi}} \int_0^{\beta_3} r^2 dr \ e^{-r^2/2} 
\end{eqnarray}
Or
\begin{eqnarray}
C = \sqrt{\frac{2}{\pi}} \left[ \int_0^{\beta_3} dr \ e^{-r^2/2} - \int_0^{\beta_3} d(r e^{-r^2/2}) \right]
\end{eqnarray}
Or
\begin{eqnarray}
C = \mathrm{erf}\left(\frac{\beta_3}{\sqrt{2}}\right) - \sqrt{\frac{2}{\pi}} \left( \beta_3 \ e^{-\beta_3^2/2} \right)
\end{eqnarray}
This transcendental equation can be solved for $\beta_3$ numerically, given the probability value $C$. Table (\ref{table:vol_scaling_factor}) shows the values of $\beta_3$ at various credible intervals $C$. Thus, the volume of the ellipsoid at confidence $C$ will be
\begin{eqnarray}
\Delta V_{C\%} = \frac{4}{3} \pi \beta_3^3 ( \sigma_x \sigma_y \sigma_z) = \frac{4}{3} \pi \beta_3^3 \sqrt{\mathrm{det}(\Sigma)}
\end{eqnarray}
where $\mathrm{det}(\Sigma)$ is the determinant of the covariance matrix $\Sigma$.

\begin{table}
\caption{\small{Ellipsoid/ellipse axes scaling factor ($\beta_q$) values at various credible intervals in $q-$dimensions}} 

\centering 
\begin{tabular}{c c c c c} 
\hline\hline 
Confidence ($C$) & $\sigma$-values & $\beta_1$ (1-D) & $\beta_2$(2-D) & $\beta_3$ (3-D)  \\ [0.5ex] 
\hline 
0.20 & 0.25$\sigma$ & 0.25 & 0.668 & 1.005  \\ 
0.683 & 1.0$\sigma$ & 1.0 & 1.516 & 1.879  \\
0.90 & 1.6$\sigma$ & 1.6 & 2.146 & 2.500  \\
0.99 & 2.6$\sigma$ & 2.6 & 3.035 & 3.368  \\ [1ex] 
\hline 
\end{tabular}
\label{table:vol_scaling_factor} 

\end{table}
In 2-dimensions, the problem is even simpler. We can start from the probability distribution in polar coordinates
\begin{eqnarray}
p(r, \theta) = \frac{1}{2 \pi} e^{-r^2/2}.
\label{eq:prob_2d_pol_symm}
\end{eqnarray}
Again let us assume that the circle of radius $\beta_2$ centered on the origin contain the probability $C$, then
\begin{eqnarray}
C = \frac{1}{2 \pi} \int_0^{\beta_2} r \ dr  \int_0^{2 \pi}  \ d\theta \ e^{-r^2/2}  = 1 - e^{\beta_2^2 / 2}
\label{eq:cir_confidence}
\end{eqnarray}
Or 
\begin{eqnarray}
\beta_2 = \sqrt{- 2 \ln(1-C)}.
\end{eqnarray}
The area of the ellipse at confidence $C$ is
\begin{eqnarray}
\Delta A_{C\%} = \pi \beta_2^2 (\sigma_x \sigma_y) = \pi \beta_2^2 \sqrt{\mathrm{det}(\Sigma)} = -2 \pi \ln(1-C) \sqrt{\mathrm{det}(\Sigma)}.
\end{eqnarray}
The values of $\beta_2$ are shown in the table (\ref{table:vol_scaling_factor}) at various confidence intervals along with $\beta_1$ for $1-$dimensional probability density which is trivial to estimate.

\end{appendix}

\end{document}